\documentclass[conference]{IEEEtran}
\IEEEoverridecommandlockouts
\usepackage{pgfplots}  
\usepackage{cite}
\usepackage{amsmath,amssymb,amsfonts}
\usepackage{graphicx}
\usepackage{textcomp}
\usepackage{xcolor}
\usepackage[noend]{algpseudocode}
\def\BibTeX{{\rm B\kern-.05em{\sc i\kern-.025em b}\kern-.08em
    T\kern-.1667em\lower.7ex\hbox{E}\kern-.125emX}}

\graphicspath{{figs/}}
\usepackage[T1]{fontenc}
\usepackage[utf8]{inputenc}
\newcommand{\ve}[1]{ {\mathbf{#1}} }
\def\RR{\mathbb{R}}
\def\bal#1{\begin{align}#1\end{align}}

\newcommand{\ds}{\ensuremath{\mathrm{s}}}
\newcommand{\dt}{\ensuremath{\mathrm{t}}}
\newcommand{\bX}{\ensuremath{\textbf{X}}}
\def\ys{\ve{y}^{\ds}}
\def\yt{\ve{y}^{\dt}}
\def\ya{\ve{y}^{\alpha}}

\newtheorem{definition}{\textbf{Definition}}

\newcommand{\R}{\ensuremath{\mathbb{R}}}

\newcommand{\bx}{\ensuremath{\textbf{x}}}

\DeclareMathOperator*{\argmin}{argmin}
\newcommand{\OT}{\text{OT}}

\begin{document}
\sloppy

\title{Audio signal interpolation using optimal transportation of spectrograms\\
\thanks{This work is supported by ANITI under the French ANR Cluster IA programme and by the France 2030 program ANR-23-PEIA-0004. Part of this work was conducted while Marien Renaud was a MSc intern student at IRIT.}
}

\author{\IEEEauthorblockN{David Valdivia}
\IEEEauthorblockA{\textit{IRIT, Universit\'e de Toulouse} \\
}
\and
\IEEEauthorblockN{Marien Renaud}
\IEEEauthorblockA{\textit{IMB, Universit\'e de Bordeaux} \\
}
\and
\IEEEauthorblockN{Elsa Cazelles, C\'edric F\'evotte}
\IEEEauthorblockA{\textit{IRIT, CNRS, Universit\'e de Toulouse}}
}

\maketitle

\begin{abstract}
We present a novel approach for generating an artificial audio signal that interpolates between given source and target sounds. Our approach relies on the computation of Wasserstein barycenters of the source and target spectrograms, followed by phase reconstruction and inversion. In contrast with previous works, our new method considers the spectrograms globally and does not operate on a temporal frame-to-frame basis. Another contribution is to endow the transportation cost matrix with a specific structure that prohibits remote displacements of energy along the time axis, and for which optimal transport is made possible by leveraging the unbalanced transport framework. The proposed cost matrix makes sense from the audio perspective and also allows to reduce the computation load. Results with synthetic musical notes and real environmental sounds illustrate the potential of our novel approach. 
\end{abstract}


\section{Introduction}

We consider the problem of generating artificial sounds that interpolate between given source and target signals. Though we consider audio signals in this paper (for creative and editing purposes), the methodology proposed in this paper is not limited to this class of signals. Given two temporal signals $\ys \in \RR^{L}$ (source) and $\yt \in \RR^{L}$ (target) we wish to generate an interpolant $\ya \in \RR^{L}$ (for any $\alpha \in [0,1]$) such that $\ve{y}^{0}=\ys $ and $\ve{y}^{1} =\yt$. To achieve this task, a baseline option is to compute the Euclidean barycenter of $\ys$ and $\yt$ with weights $(1-\alpha,\alpha)$ given by
\bal{
\ve{y}^{\alpha} &= \argmin_{\ve{y}} \ (1- \alpha) \, \| \ve{y} - \ve{y}^{\rm s} \|^{2}_{2}  + \alpha \, \| \ve{y} - \ve{y}^{\rm t} \|^{2}_{2} \label{eq:eucbar} \\
 & = (1-\alpha) \, \ve{y}^{\rm s} + \alpha \, \ve{y}^{\rm t} \label{eq:eucbar2}.
}
As shown by \eqref{eq:eucbar2}, the Euclidean barycenter is just a weighted average of the source and target signals. The objective of this paper is to explore an alternative option based on {\em optimal transport} (OT), that can produce truly hybrid sounds. In the proposed setting, the normalized time-frequency (t-f) distributions of the source and target signals are interpreted as discrete probability distributions supported by the t-f grid. We then propose to compute a Wasserstein barycenter \cite{Agueh2011}, also called an interpolant, of these t-f distributions. The resulting t-f barycenter may then be inverted using suitable t-f grid reassignment and phase reconstruction to obtain a temporal interpolant. We also propose to endow the transportation cost matrix with a specific structure that prohibits remote displacements of energy along the time axis, and for which optimal transport is made possible by leveraging the unbalanced transport framework. The proposed cost matrix makes sense from the audio perspective and also allows to reduce the computation load.

To the best of our knowledge, \cite{Henderson2019} is the first prior work to consider optimal transport of spectra for audio interpolation. The setting in \cite{Henderson2019} is however more restricted than ours because the authors consider the frames of the spectrogram individually (frame-to-frame OT). Furthermore, the source and target t-f  distributions result from a specific pitch-based processing of the spectrograms, which exploits the harmonic nature of tonal musical signals. Peak frequencies are identified in the source and target spectra using t-f reassignment \cite{Auger2013}. Those peak frequencies and their energy (integrated over a selected neighborhood) lay ground to the OT procedure. This approach is well suited to pitched sounds. Another prior work in audio interpolation with OT is \cite{Roma2020}. In this work, a nonnegative matrix factorization (NMF) of the source and target spectrograms is first computed \cite{Smaragdis2014}. Simplifying a little, an interpolated spectrogram is then constructed by multiplying the temporal activations of the target spectrogram with a dictionary of interpolated spectra. The latter are obtained by interpolating the source and target spectral dictionaries obtained by NMF, using the OT procedure of \cite{Henderson2019}. In contrast with the approach of \cite{Henderson2019,Roma2020}, we consider the spectrogram {\em globally}, meaning that transportation of energy is possible over neighboring time frames. Furthermore, we do not resort to any pre-processing and work directly on the raw spectrograms. This allows to work with a wider range of sounds, and in particular wide-band signals. Other technical differences related to the support of the interpolated distribution and to phase reconstruction will be discussed next.

While OT has been considered in many image processing problems, it has been less popularized in audio signal processing settings and this paper is also a step in that direction. Besides \cite{Henderson2019,Roma2020} that consider audio interpolation, other applicative examples of OT in audio signal processing are music transcription \cite{Flamary2016,Elvander2017} and audio classification \cite{Dessein2018,Cazelles2021}. The methods derived in this paper are non-supervised (the source and target signals are the only two inputs). Supervised learning-based approaches have also been considered for audio interpolation, see, e.g., \cite{Engel2019}. These approaches can produce spectacular results but rely on the intensive training of generative networks.\footnote{Audio interpolation may also refer to the tasks of filling missing gaps in a corrupted signal \cite{Godsill1998,Adler2012} or synthesizing virtual 3D sounds \cite{Radke2002}. Such tasks are not considered here but could benefit from our results in future work.}

{The paper is organized as follows. Section \ref{sec:preliminaries} introduces notations and elements of OT. We then present a first interpolation method based on Wasserstein barycenters in Section \ref{sec:method}. Some limitations of this baseline method are remedied by our second method, that relies on unbalanced OT and a structured cost matrix, as described in Section~\ref{sec:uot}. Finally, we report experiments with synthetic musical notes and real environmental sounds in Section \ref{sec:experiments}, before concluding in Section \ref{sec:conclusion}.}


\section{Preliminaries}\label{sec:preliminaries}

\subsection{Notations}

We denote by $\bX^{\ds}$, $\bX^{\dt} \in \RR_{+}^{M \times N}$ the spectrograms of the source and target signals. They are equal to the magnitude of a short-time Fourier transform (STFT) of $\ve{y}^{\ds}$ and $\ve{y}^{\dt}$. The values of $M$ (number of frequency bins) and $N$ (number of time frames) are dictated by the length of the chosen STFT analysis window and hop size. Dropping the superscripts $\ds$ and $\dt$ for conciseness, the coefficients of the spectrogram $\bX$ are denoted by $X_{mn} \in \RR_{+}$, with $(m,n) \in \{1,\ldots, M\} \times \{1,\ldots, N\} $. The t-f couples $(m,n)$ index t-f points of the form $\omega_{mn} = (f_{m}, t_{n})$ with $f_{m} = \frac{m-1}{M} \frac{f_{s}}{2} $ and $t_{n} = (n-1)\frac{H}{f_{s}}$, where $f_{s}$ and $H$ denote the sampling frequency (Hz) and hop size (in samples).

Let us define by $\ve{x}^{\ds}$ and $\ve{x}^{\dt}$ the vectorized versions of $\bX^{\ds}$ and $\bX^{\dt}$. The vectorization underlies an arbitrary one-to-one mapping between the t-f grid $(m,n) \in \{1,\ldots, M\} \times \{1,\ldots, N\} $ and a point $i \in \{1,\dots,I\}$, with $I = MN$, such that $x^{\ds}_{i} =  X^{\ds}_{mn}$ and $x^{\dt}_{i} =  X^{\dt}_{mn}$. We will sometimes use the shorthand $\omega_{i} = \omega_{mn} $. 
%
%
We assume that the spectrograms have been normalized (globally), so that $\sum_{i} x^{\ds}_{i} = \sum_{i} x^{\dt}_{i} =1 $. 

\subsection{Optimal transport and Wasserstein barycenters} \label{sec:OT}

Equipped with these notations and conventions, we may view the source and target spectrograms $\bX^{\ds}$ and $\bX^{\dt}$ as discrete probability distributions $\mu^{\ds}$ and $\mu^{\dt}$ supported by the t-f points $\omega_{i}\in\R^2$, i.e.,
\begin{equation}\label{eq:tf_distribution}
\mu = \sum_{m=1}^M\sum_{n=1}^N X_{mn} \delta_{\omega_{mn}} = \sum_{i=1}^I x_{i} \delta_{\omega_{i}}
\end{equation}
where we dropped the superscripts $\ds$ or $\dt$, and where $\delta$ is the Dirac delta function. Using OT terminology, the coefficients $X_{mn}$ or $x_i$ will sometimes be referred to as ``mass".

\begin{definition} \label{Definition_1}
The optimal transport distance, or 2-Wasserstein distance, between distributions $\mu^{\ds}$ and $\mu^{\dt}$ is given by $\sqrt{\OT \left( \mu^{\ds},\mu^{\dt} \right)}$, with
\begin{equation}
\label{eq:wass}
\OT \left( \mu^{\ds},\mu^{\dt} \right) = \min_{\textbf{P} \in \Pi(\bx^{\ds},\bx^{\dt})}{\langle \textbf{C},\textbf{P} \rangle },
\end{equation}
where $\Pi(\bx^{\ds},\bx^{\dt}):=\{ \textbf{P} \in \R^{I\times I}_{+} | \textbf{P} \textbf{1}_{I} = \bx^{\ds}, \textbf{P}^{\top}\textbf{1}_I = \bx^{\dt} \}$ is the set of transport plans $\ve{P}$ with marginals $\bx^{\ds}$ and $\bx^{\dt}$ ($\textbf{1}_I$ denotes the vector in $\R^I$ with all entries equal to one), and $\textbf{C}$ is the transportation cost matrix defined entry-wise by $C_{ii'} = \Vert\omega_i-\omega_{i'}\Vert^2$.
\end{definition}

\begin{definition}
The Wasserstein barycenter of coefficient $\alpha \in [0,1]$ between the two distributions $\mu^{\ds}$ and $\mu^{\dt}$ is defined by
\begin{equation}
\label{eq:wbar}
\mu^{\alpha} \in \argmin_{\nu} \ (1-\alpha)\, \OT\left(\mu^{\ds}, \nu \right)^2 + \alpha \, \OT\left(\mu^{\dt},\nu\right)^2. 
\end{equation}
\end{definition}
It can easily be shown with the triangular inequality that $\mu^{\alpha}$ is explicit in terms of an optimal plan $\ve{P}^\star\in\Pi(\bx^{\ds},\bx^{\dt})$ in \eqref{eq:wass}  \cite{peyre2019computational}, and is given by 
\begin{equation}
\label{eq:explicit_bar}
\mu^{\alpha} = \sum_{i,i'=1}^I P^{\star}_{ii'} \, \delta_{(1-\alpha)\omega_i+\alpha \omega_{i'}}.
\end{equation}
Note that $\ve{P}^\star$ is independent of $\alpha$.

\section{Signal interpolation using time-frequency Wasserstein barycenters} \label{sec:method}

When $\mu^{\ds}$ and $\mu^{\dt}$ are the t-f distributions \eqref{eq:tf_distribution} of the source and target signals, the distribution $\mu^{\alpha}$ given by \eqref{eq:explicit_bar} defines a t-f barycenter that one might wish to invert back into the time domain in order to reconstruct an audio interpolant $\ve{y}^{\alpha}$ of $\ve{y}^{\ds}$ and $\ve{y}^{\dt}$. This simple and yet novel idea raises practical issues regarding the t-f support of $\mu^{\alpha}$ and phase reconstruction, as discussed next.

\subsection{Time-frequency reassignment} As indicated by \eqref{eq:explicit_bar}, the support of $\mu^{\alpha}$ is the set of t-f points $\omega_{ii'}^{\alpha} := (1-\alpha) \omega_{i} + \alpha \omega_{i'}$, $(i, i') \in \{ 1, \ldots, I \}^{2}$.
Equivalently, given $\alpha \in [0,1]$, the support of $\mu^{\alpha}$ is given by the $(MN)^{2}$ t-f points $\{ \left( (1-\alpha) f_{m} + \alpha f_{m'},(1-\alpha) t_{n} + \alpha t_{n'}   \right)\}_{mm'nn'}$.  As such, $\mu^{\alpha}$ is not supported by the same t-f grid than $\mu^{\ds}$ and $\mu^{\dt}$, it instead is spread over many more t-f points which form an irregular grid. In this paper, we choose to reconstruct a native-grid spectrogram $\ve{X}^{\alpha} \in \RR_{+}^{M \times N} $ from $\mu^{\alpha}$ by reassigning the mass $P^{\star}_{ii'}$ supported by $\omega_{ii'}^{\alpha}$ to the closest native t-f point $\omega_{j^{\star}}$, such that $j^{\star} = \argmin_{j} \| \omega_{j} - \omega_{ii'}^{\alpha} \| $.

\subsection{Phase reconstruction} Given the barycentric spectrogram $\ve{X}^{\alpha} $ reconstructed in the previous step, we use the classical Griffin and Lim phase reconstruction algorithm \cite{Griffin1983} and overlap-add inverse STFT to reconstruct a temporal signal $\ve{y}^{\alpha}$.

\subsection{Discussion}

{The baseline approach presented at the beginning of this section presents some limitations that we now discuss. Firstly, the computation of $\ve{P}^{\star}$ in \eqref{eq:wass} might be too computationally intensive even for small-size problems corresponding to few seconds of audio. A standard remedy is to consider entropy-regularized OT, which can be solved more efficiently using the Sinkhorn algorithm \cite{Cuturi13}. Unfortunately, we found out that this solution is not desirable for spectrogram interpolation, as the entropy term tends to spread the mass of the optimal plan too smoothly, cutting out time-frequency details. Secondly, even for short audio signals, we found preferable to prevent the mass from moving to anywhere on the t-f support. As such, we propose in the next section a structured cost matrix that favors mass displacement along the frequency axis while controlling the permissible displacement over neighboring time frames. We then leverage the unbalanced OT framework to efficiently compute the corresponding optimal transport plan.}


\section{Signal interpolation using unbalanced optimal transport and a structured cost matrix} \label{sec:uot}

\subsection{Structured cost matrix}\label{sec:cost_matrix}

{In order to produce audio results that sound more natural}, we propose to prevent remote time displacements of mass by introducing the cost matrix $\bar{\ve{C}} \in \RR_+^{I \times I}$ with coefficients
\begin{equation}
    \label{eq:costC}
    \bar{C}_{ii'} = \left\{
    \begin{array}{ll}
        (f_m - f_{m'})^2 +
        (t_{n} - t_{n'})^2  & \mbox{if } |n - n'|\leq p, \\
        +\infty & \mbox{otherwise},
    \end{array}
\right.
\end{equation}
where $(f_m,t_n)$ (resp. $(f_{m'},t_{n'})$) is the t-f point mapped with $i$ (resp. $i'$). In contrast with the standard $\ell_2^2$ cost matrix $\ve{C}$ introduced in Section~\ref{sec:OT}, $\bar{\ve{C}}$ now explicitly forbids mass displacements of time range larger than $p$ frames. As our experiments will show, the user-defined parameter $p$ plays a significant role.\footnote{In practice, we use dimensionless time and frequency variables and simply set $C_{ii'} = (m-m')^2 + (n-n')^2$, and likewise for $\bar{C}_{ii'}$.} Note that when $p=0$, the cost matrix $\bar{\ve{C}}$ only allows frequency displacements in matching frames $n=n'$ (mass cannot spill over adjacent frames). This is close to the frame-to-frame setting of \cite{Henderson2019}, with the exception that in our case the spectrograms are still considered as a whole distribution (and normalized globally) as opposed to every frame being normalized like in \cite{Henderson2019}.


\subsection{Unbalanced OT and spectrogram interpolants}


As the proposed structured cost matrix \eqref{eq:costC} proscribes certain displacement of mass, the OT problem \eqref{eq:wass} using $\bar{\ve{C}}$ might not have a solution verifying the marginals constraint \cite{peyre2019computational}. We therefore consider unbalanced OT \cite{chizat2018unbalanced,liero2018optimal} which relaxes the conservation of mass constraints in problem \eqref{eq:wass}:
\begin{align}
\text{UOT}_\beta \left( \mu^{\ds},\mu^{\dt} \right) = \min_{\textbf{P} \in \R_+^{I\times I}}{\langle\textbf{C},\textbf{P} \rangle } &+ \beta \ \text{KL}(\textbf{P}\mathbf{1}_I, \bx^{\ds}) \nonumber\\
&+ \beta \ \text{KL}(\textbf{P}^T \mathbf{1}_I, \bx^{\dt}),
\label{eq:unbalanced_wass}
\end{align}
where $\text{KL}$ is the Kullback-Leibler divergence between nonnegative numbers, and $\beta\in\R_+$ is a parameter that controls the trade-off between mass transportation and mass creation/loss. For a large $\beta$, one recovers the OT problem \eqref{eq:wass}, whereas a small $\beta$ will encourage displacements of mass between points that are close in the t-f grid, promoting a local consistency of frequency content between the source and target spectrograms.

Following \eqref{eq:explicit_bar}, we reconstruct a t-f interpolant from an optimal solution $\ve{P}^{\beta}\in\R_+^{I\times I}$ of \eqref{eq:unbalanced_wass}:
\begin{equation}\label{eq:UOT_bar}
    \mu^{\alpha}_\beta = \sum_{i,i'=1}^I P^{\beta}_{ii'} \, \delta_{(1-\alpha)\omega_i+\alpha \omega_{i'}}.
\end{equation}
An audio interpolant $\ve{y}^{\alpha}$ is then reconstructed from $\mu^{\alpha}_\beta$ as in Section \ref{sec:method}.


\section{Experiments}\label{sec:experiments}

\subsection{Implementation details and experimental setting}

The interpolant \eqref{eq:explicit_bar} requires to solve the OT problem \eqref{eq:wass}, for which we used the POT toolbox \cite{flamary2021pot}. The interpolant \eqref{eq:UOT_bar} requires to solve the UOT problem \eqref{eq:unbalanced_wass}, for which we use the majorization-minimization (MM) algorithm introduced in \cite{chapel2021unbalanced}. We implemented a dedicated version of the MM algorithm that leverages the particular structure of $\bar{\ve{C}}$ in \eqref{eq:costC}. Indeed, we may only consider indices $ii'$ with finite entries $\bar{C}_{ii'}$, because $P_{ii'}^\beta= 0$ whenever $\bar{C}_{ii'}= +\infty$. This allows for an efficient implementation that can run with audio signals of up to a few seconds with standard sampling frequencies.

In the following, we present results with synthetic musical notes and real environmental sounds. For comparison with OT and readability of the spectrograms, we use short audio signals of duration 1 second sampled at $f_s=$16kHz. The spectrograms are computed with a Hann window of duration 40ms with overlap 50\%. Audio samples and code are available online.\footnote{https://davidvaldiviad.github.io/audio-signal-interpolation-ot/}



\subsection{Interpolating musical notes}

\begin{figure}[t]
\footnotesize
  \centering
  \begin{tabular}{c} 
    \begin{tikzpicture}
      \begin{axis}[
        width=0.8\columnwidth,
        height=0.25\columnwidth,
        ylabel={Frequency (kHz)},
        scale only axis,
        title={(a) Piano (source)},
        xtick={0,0.2,0.4,0.6,0.8,1},
        ytick={0,2,4,6,8},
        xmin=0, xmax=1,
        ymin=0, ymax=8,
        xlabel near ticks,
        ylabel near ticks,
        tick label style={font=\scriptsize},
        label style={font=\scriptsize},
      ]
        \addplot graphics[xmin=-0.02,xmax=1.01,ymin=-0.2,ymax=8.2] {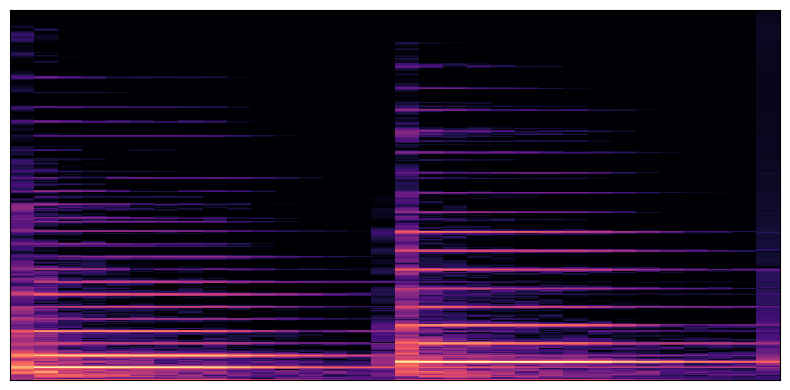};
      \end{axis}
    \end{tikzpicture}
    \\
    \begin{tikzpicture}
      \begin{axis}[
        width=0.8\columnwidth,
        height=0.25\columnwidth,
        ylabel={Frequency (kHz)},
        scale only axis,
        title={(b) Guitar (target)},
        xtick={0,0.2,0.4,0.6,0.8,1},
        ytick={0,2,4,6,8},
        xmin=0, xmax=1,
        ymin=0, ymax=8,
        xlabel near ticks,
        ylabel near ticks,
        tick label style={font=\scriptsize},
        label style={font=\scriptsize},
      ]
        \addplot graphics[xmin=-0.02,xmax=1.01,ymin=-0.2,ymax=8.2] {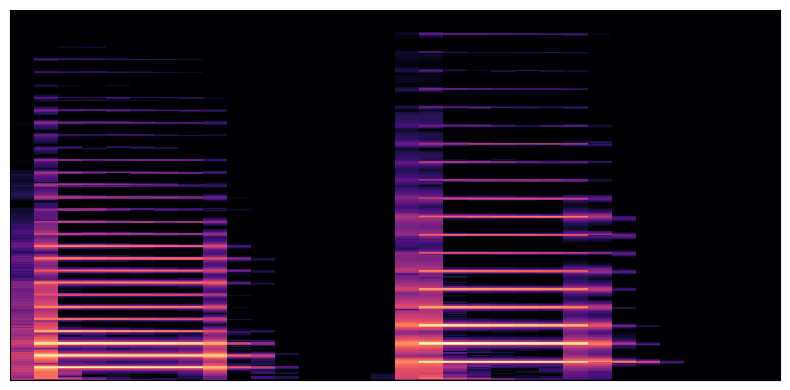};
      \end{axis}
    \end{tikzpicture}
    \\
    \begin{tikzpicture}
      \begin{axis}[
        width=0.8\columnwidth,
        height=0.25\columnwidth,
        ylabel={Frequency (kHz)},
        scale only axis,
        title={(c) OT interpolation},
        xtick={0,0.2,0.4,0.6,0.8,1},
        ytick={0,2,4,6,8},
        xmin=0, xmax=1,
        ymin=0, ymax=8,
        xlabel near ticks,
        ylabel near ticks,
        tick label style={font=\scriptsize},
        label style={font=\scriptsize},
      ]
        \addplot graphics[xmin=-0.02,xmax=1.01,ymin=-0.2,ymax=8.2] {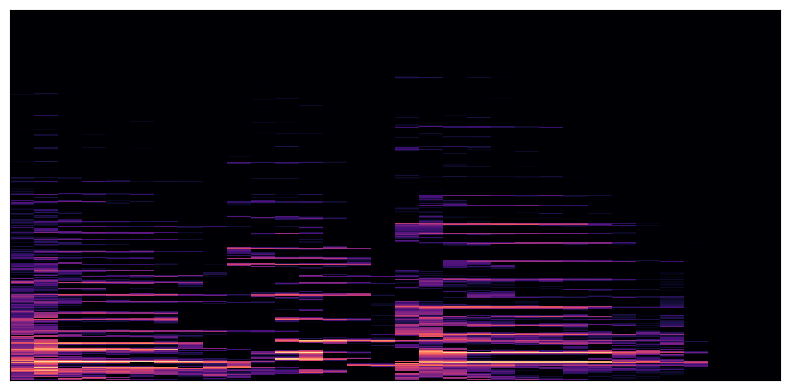};
      \end{axis}
    \end{tikzpicture}
    \\
    \begin{tikzpicture}
      \begin{axis}[
        width=0.8\columnwidth,
        height=0.25\columnwidth,
        xlabel={Time (s)},
        ylabel={Frequency (kHz)},
        scale only axis,
        title={(d) UOT interpolation ($p=0$)},
        xtick={0,0.2,0.4,0.6,0.8,1},
        ytick={0,2,4,6,8},
        xmin=0, xmax=1,
        ymin=0, ymax=8,
        xlabel near ticks,
        ylabel near ticks,
        tick label style={font=\scriptsize},
        label style={font=\scriptsize},
      ]
        \addplot graphics[xmin=-0.02,xmax=1.01,ymin=-0.2,ymax=8.2] {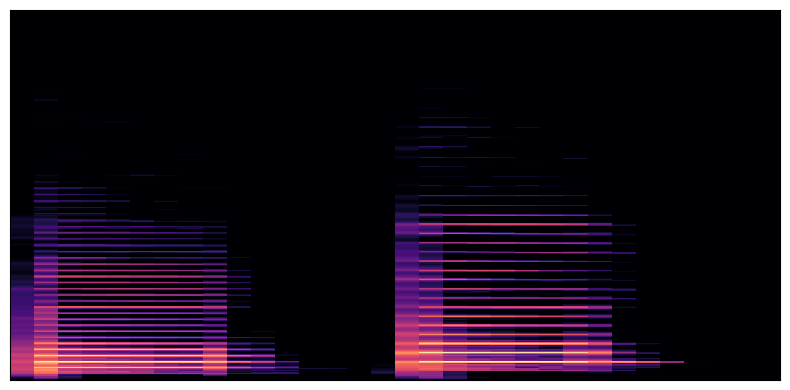};
      \end{axis}
    \end{tikzpicture}
  \end{tabular}
\caption{Interpolation between two versions of the same sequence of notes C3-G3 played on a piano and a guitar, using $\alpha=0.5$.}
  
  \label{fig:melodies-interpolation}
\end{figure}

Our first experiment consists in interpolating two versions of the same sequence C3-G3 played either with a piano (source) or a guitar (target). The two input signals were generated with the music production software Ableton Live. We set the interpolation parameter to $\alpha=0.5$. For the UOT-based method, we set the hyperparameter to $\beta=1$ and the time-limiting parameter to $p=0$. Results are shown in Fig. \ref{fig:melodies-interpolation}.

We observe that the structure of $\bar{\ve{C}}$ has a decisive influence on the nature of the interpolants. Since OT is {\em constrained} to move all the mass from the source to the target, some mass is spread along the time axis. This produces a continuous audio signal in which the temporal structure of the source and target signals is lost, see Fig. \ref{fig:melodies-interpolation}(c). On the other hand, by limiting the mass displacement on the time axis, UOT is able to produce an interpolant that preserves the dynamics of the input signals as seen in Fig. \ref{fig:melodies-interpolation}(d). This results in the same sequence C3-G3 played by a hybrid instrument (in between a guitar and a piano), as can be gathered from the online audio samples.


\subsection{Interpolating environmental sounds}

We now consider environmental sounds that have a more {\em textured} structure as opposed to mostly harmonic. The source is the chirp of a cicada and the target is a sample of flowing water. The cicada's chirp is characterized by strong high-frequency energy and by a clear rhythmic pattern, whereas the water flow is continuous and essentially wide-band.

\begin{figure}[t]
\footnotesize
  \centering
  \begin{tabular}{c} 
    \begin{tikzpicture}
      \begin{axis}[
        width=0.8\columnwidth,
        height=0.25\columnwidth,
        ylabel={Frequency (kHz)},
        scale only axis,
        title={(a) Cicada's chirp (source)},
        xtick={0,0.2,0.4,0.6,0.8,1},
        ytick={0,2,4,6,8},
        xmin=0, xmax=1,
        ymin=0, ymax=8,
        xlabel near ticks,
        ylabel near ticks,
        tick label style={font=\scriptsize},
        label style={font=\scriptsize},
      ]
        \addplot graphics[xmin=-0.02,xmax=1.01,ymin=-0.2,ymax=8.2] {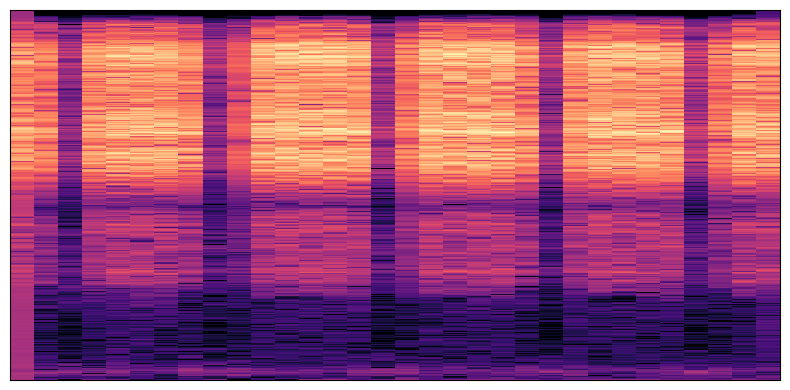};
      \end{axis}
    \end{tikzpicture}
    \\
    \begin{tikzpicture}
      \begin{axis}[
        width=0.8\columnwidth,
        height=0.25\columnwidth,
        ylabel={Frequency (kHz)},
        scale only axis,
        title={(b) Flowing water (target)},
        xtick={0,0.2,0.4,0.6,0.8,1},
        ytick={0,2,4,6,8},
        xmin=0, xmax=1,
        ymin=0, ymax=8,
        xlabel near ticks,
        ylabel near ticks,
        tick label style={font=\scriptsize},
        label style={font=\scriptsize},
      ]
        \addplot graphics[xmin=-0.02,xmax=1.01,ymin=-0.2,ymax=8.2] {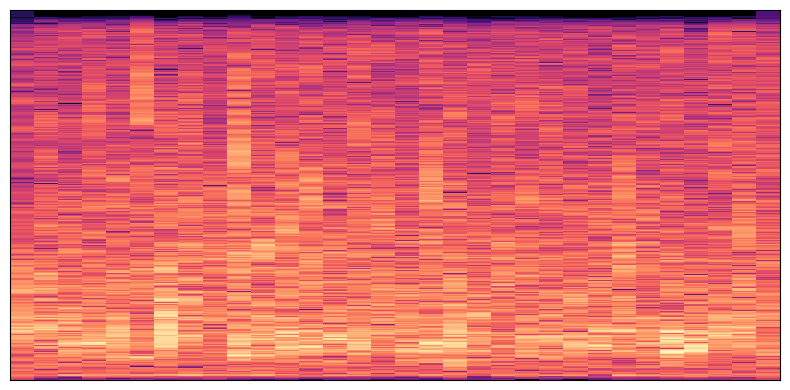};
      \end{axis}
    \end{tikzpicture}
    \\
    \begin{tikzpicture}
      \begin{axis}[
        width=0.8\columnwidth,
        height=0.25\columnwidth,
        ylabel={Frequency (kHz)},
        scale only axis,
        title={(c) OT interpolation},
        xtick={0,0.2,0.4,0.6,0.8,1},
        ytick={0,2,4,6,8},
        xmin=0, xmax=1,
        ymin=0, ymax=8,
        xlabel near ticks,
        ylabel near ticks,
        tick label style={font=\scriptsize},
        label style={font=\scriptsize},
      ]
        \addplot graphics[xmin=-0.02,xmax=1.01,ymin=-0.2,ymax=8.2] {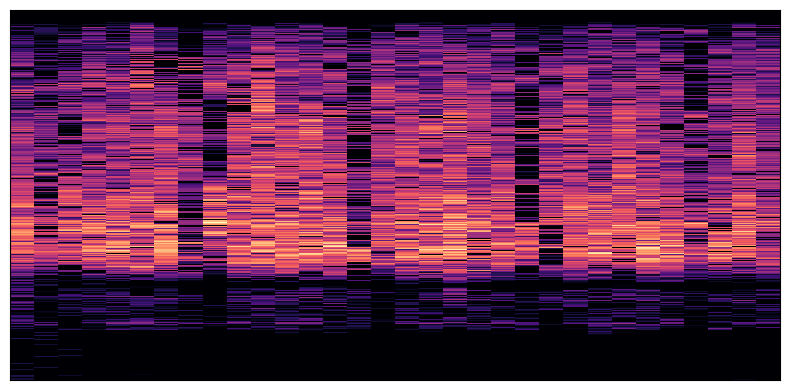};
      \end{axis}
    \end{tikzpicture}
    \\
    \begin{tikzpicture}
      \begin{axis}[
        width=0.8\columnwidth,
        height=0.25\columnwidth,
        ylabel={Frequency (kHz)},
        scale only axis,
        title={(d) UOT interpolation ($p=0$)},
        xtick={0,0.2,0.4,0.6,0.8,1},
        ytick={0,2,4,6,8},
        xmin=0, xmax=1,
        ymin=0, ymax=8,
        xlabel near ticks,
        ylabel near ticks,
        tick label style={font=\scriptsize},
        label style={font=\scriptsize},
      ]
        \addplot graphics[xmin=-0.02,xmax=1.01,ymin=-0.2,ymax=8.2] {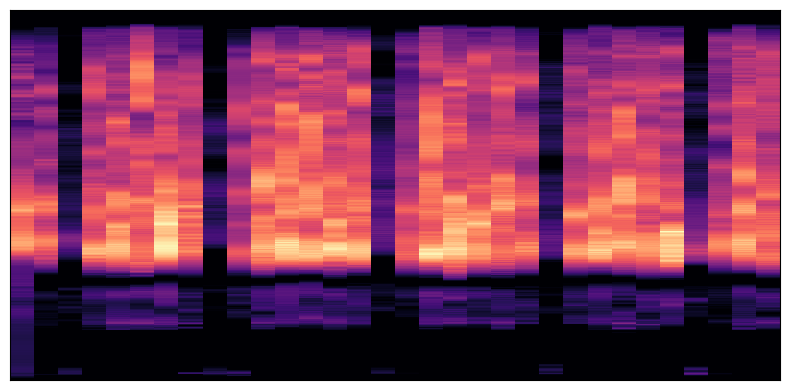};
      \end{axis}
    \end{tikzpicture}
    \\
    \begin{tikzpicture}
      \begin{axis}[
        width=0.8\columnwidth,
        height=0.25\columnwidth,
        ylabel={Frequency (kHz)},
        scale only axis,
        title={(e) UOT interpolation ($p=5$)},
        xtick={0,0.2,0.4,0.6,0.8,1},
        ytick={0,2,4,6,8},
        xmin=0, xmax=1,
        ymin=0, ymax=8,
        xlabel near ticks,
        ylabel near ticks,
        tick label style={font=\scriptsize},
        label style={font=\scriptsize},
      ]
        \addplot graphics[xmin=-0.02,xmax=1.01,ymin=-0.2,ymax=8.2] {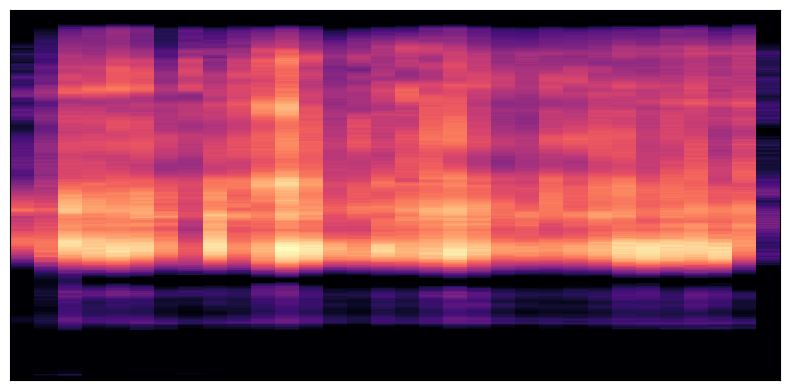};
      \end{axis}
    \end{tikzpicture}
    \\
    \begin{tikzpicture}
      \begin{axis}[
        width=0.8\columnwidth,
        height=0.25\columnwidth,
        xlabel={Time (s)},
        ylabel={Frequency (kHz)},
        scale only axis,
        title={(f) UOT interpolation ($p=+\infty$)},
        xtick={0,0.2,0.4,0.6,0.8,1},
        ytick={0,2,4,6,8},
        xmin=0, xmax=1,
        ymin=0, ymax=8,
        xlabel near ticks,
        ylabel near ticks,
        tick label style={font=\scriptsize},
        label style={font=\scriptsize},
      ]
        \addplot graphics[xmin=-0.02,xmax=1.01,ymin=-0.2,ymax=8.2] {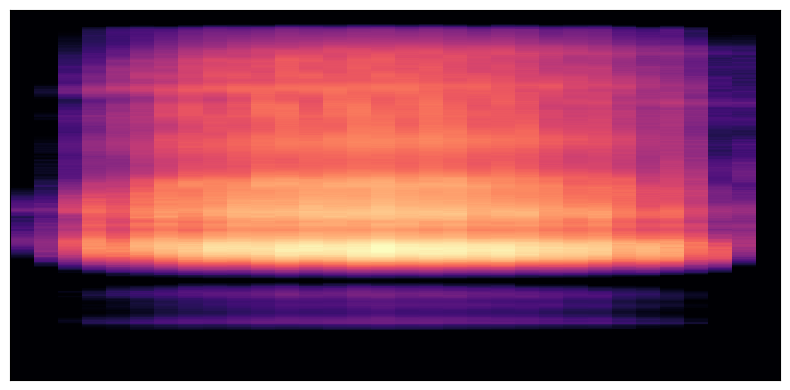};
      \end{axis}
    \end{tikzpicture}
  \end{tabular}

\caption{Interpolation between a cicada's chirp and flowing water, using $\alpha=0.5$. Various values of the time-limiting parameter $p$ are considered. In subplot (f) we use $p=+\infty$ as a shorthand for using $\bar{\ve{C}}= \ve{C}$ (the time limit is lifted).}

  \label{fig:texture-interpolation}
\end{figure}

We set $\alpha=0.5$ and $\beta=1$ like in previous section, but we now explore various values of the time-limiting parameter $p$, see Fig.~\ref{fig:texture-interpolation}. Fig.~\ref{fig:texture-interpolation}(c) and Fig.~\ref{fig:texture-interpolation}(d) show that OT and UOT with $p=0$ capture essential characteristics of the input signals: the temporal structure induced by the cicada's chirp and wide-band energy content. However, Fig.~\ref{fig:texture-interpolation}(c) shows that the temporal structure of the OT interpolant is not as sharp, again because of the conservation of mass constraint. Increasing the value of $p$ for the UOT interpolant has a significant impact, see Figs.~\ref{fig:texture-interpolation}(e-f): larger values of $p$ allow for increasing temporal transportation (i.e., along the temporal axis). The perceptual impact of the value of $p$ can be assessed by listening to the audio samples available online.



\section{Discussion}\label{sec:conclusion}

In our opinion, our work opens promising research directions regarding the optimal transportation of spectrograms for signal processing applications. In this paper we have resorted to a t-f grid reassignment followed by standard phase retrieval to proceed to the temporal reconstruction of the interpolants. An exciting research direction would be to design inversion methods that circumvent the reassignment and work directly in the irregular grid. Another direction would be to design t-f interpolants that are constrained to be supported by the native grid. A third direction would be to leverage the phase of the source and target signals to produce better estimates of the phase of the interpolant. This latter direction in particular may further improve audio quality. 



\clearpage


\end{document}